\documentclass[twocolumn,pra,showpacs,superscriptaddress]{revtex4}
\usepackage{amssymb}
\usepackage{amsmath}
\usepackage{graphicx}
\usepackage{subfigure}
\usepackage{natbib}
\usepackage{epsfig}
\usepackage{amsfonts}
\usepackage{mathrsfs}
\usepackage{CJK}
\usepackage[toc,page,title,titletoc,header]{appendix}

\begin{document}
\title{Quantum speed limits for Bell-diagonal states}

\author{Yingjie Zhang }
\email{qfyingjie@iphy.ac.cn}
 \affiliation{Shandong Provincial Key
Laboratory of Laser Polarization and Information Technology,
Department of Physics, Qufu Normal University, Qufu 273165, China}
 \affiliation{Institute of
Physics, Chinese Academy of Sciences, Beijing, 100190, China}

\author{Wei Han}
 \affiliation{Shandong Provincial Key
Laboratory of Laser Polarization and Information Technology,
Department of Physics, Qufu Normal University, Qufu 273165, China}

\author{Yunjie Xia }
\email{yjxia@mail.qfnu.edu.cn}
 \affiliation{Shandong Provincial Key
Laboratory of Laser Polarization and Information Technology,
Department of Physics, Qufu Normal University, Qufu 273165, China}

\author{Kexia Jiang }
 \affiliation{Institute of Physics,
Chinese Academy of Sciences, Beijing, 100190, China}
\author{Junpeng Cao }
 \affiliation{Institute of Physics,
Chinese Academy of Sciences, Beijing, 100190, China}
\author{Heng Fan }
\email{hfan@iphy.ac.cn}
 \affiliation{Institute of Physics,
Chinese Academy of Sciences, Beijing, 100190, China}

\date{\today}
\begin{abstract}
Bounds of the minimum evolution time between two distinguishable
states of a system can help to assess the maximal speed of quantum
computers and communication channels. We study the quantum speed
limit time of a composite quantum states in the presence of
nondissipative decoherence. For the initial states with maximally
mixed marginals, we obtain the exactly expressions of quantum speed
limit time which mainly depend on the parameters of the initial
states and the decoherence channels. Furthermore, by calculating
quantum speed limit time for the time-dependent states started from
a class of initial states, we discover that the quantum speed limit
time gradually decreases in time, and the decay rate of the quantum
speed limit time would show a sudden change at a certain critical
time. Interestingly, at the same critical time, the composite system
dynamics would exhibit a sudden transition from classical to quantum
decoherence.

\end{abstract}

\maketitle

\section{\textbf{{Introduction}}}
Quantum mechanics establishes the fundamental bounds for the minimum
evolution time between two states of a given system. Bounds of this
evolution time, known as quantum speed limit time (QSLT), are
intimately related to the evolution speed of quantum systems. The
applications of these bounds are shown in different scenarios,
including quantum communication \cite{1}, quantum metrology
\cite{2}, quantum computation \cite{3}, as well as quantum optimal
control protocols \cite{4}. Derivations of these basic bounds
focused on the closed system with unitary evolution, are obtained by
Mandelstam-Tamm (MT) type bound and Margolus-Levitin (ML) type bound
\cite{5,6,7,8,9,10,a1,a2,a3,a4,a5,a6}. Moreover, the extensions of
the QSLT for nonunitary evolution of open systems are also studied
\cite{11,12,13,14}, since the relevant influence of the environment
on processing or information transferring systems can not be
ignored. It is shown that two unified bounds of QSLT for the pure or
mixed initial states including both MT and ML types for nonunitary
dynamics process have been formulated, respectively \cite{13,14}.
Although these two unified bounds have already been used to
illustrate the quantum evolution speed for a qubit state (whether it
is a pure state or mixed state) under decoherence process. As we all
know that the correlated composite quantum states are playing a
central role in quantum information and communication theory
\cite{15,b1}. Such as, in order to fulfil a long-distance
high-fidelity quantum communication in a noisy channel, entanglement
of a composite system (two distant sites) must be used to implement
quantum repeaters \cite{b2}. And the QSLT on the evolution of an
open composite quantum system would help to deal with the robustness
of quantum simulators and computers against decoherence \cite{b3}.
So the application of these unified bounds to assess the evolution
speed of the arbitrary states in a composite quantum system with
nonunitary dynamics, is very much looking forward.

In this paper, we are interested in the evolution speed of two-qubit
Bell-diagonal states under certain noise channels (i.e., phase flip,
bit flip, and bit-phase flip). By making use of the unified bound of
QSLT for arbitrary mixed states \cite{14}, we have identified the
conditions on the different expressions of the QSLT for the initial
two-qubit Bell-diagonal states under decoherence process. In the
following, we also focus on the QSLT for the time-dependent states
which started from the initial Bell-diagonal states in the whole
dynamics process. We demonstrate that the QSLT will be reduced
gradually with the starting point in time under the noise channels,
that is to say, the two-qubit system executes a speeded-up dynamics
evolution process. Moreover, for certain initial Bell-diagonal
states considered in Ref. \cite{16}, a remarkable result we find in
the paper is the existence of a sudden change of the decay rate of
the QSLT at a certain critical time $\tau_{c}$. And at the same
critical time $\tau_{c}$, a sudden transition from classical to
quantum decoherence for this class of initial Bell-diagonal states
also appears. Then we can specifically point out that the evolution
speed of the whole dynamics process can be described qualitatively
by the classical decoherence for $\tau<\tau_{c}$, and quantum
decoherence for $\tau>\tau_{c}$. Finally, we also recognize a
symmetry among the above three decoherence channels.

 This paper is organized as follows. In Sec.
II, we first review the definition of the QSLT, and we give the
explicit formulas of the QSLT for the two-qubit initial
Bell-diagonal states under certain noise channels. In Sec. III, we
apply the QSLT of the time-dependent states to illustrate the speed
of the whole dynamics evolutionary process. We conclude in the last
section.
\section{\textbf{{Quantum speed limit under decoherence channels}}}
 Firstly, we start with the definition of the QSLT for
open quantum systems. A unified lower bound, including both MT and
ML types, has been derived in Refs. \cite{13,14}. With the help of
the von Neumann trace inequality and the Cauchy-Schwarz inequality,
the QSLT between an arbitrary initially mixed state $\rho_{0}$ and
its target state $\rho_{\tau_{D}}$, governed by the master equation
$\dot{\rho}_{t}=L_{t}\rho_{t}$, with $L_{t}$ the positive generator
of the dynamical semigroup, is as follows
\begin{equation}
\tau_{QSL}=\max\{\frac{1}{\overline{\sum^{n}_{i=1}\sigma_{i}\varrho_{i}}},\frac{1}{\overline{\sqrt{\sum^{n}_{i=1}\sigma^{2}_{i}}}}\}B(\rho_{0},\rho_{\tau_{D}}),
\label{1}
\end{equation}
with $\overline{X}=\tau_{D}^{-1}\int^{\tau_{D}}_{0}Xdt$,
$B(\rho_{0},\rho_{\tau_{D}})=|tr(\rho_{0}\rho_{\tau_{D}})-tr(\rho^{2}_{0})|$
denotes the Bures distance between the initial state $\rho_{0}$ and
the target state $\rho_{\tau_{D}}$, and $\sigma_{i}$ are the
singular values of $\dot{\rho}_{t}$ and $\varrho_{i}$ those of the
initial mixed state $\rho_{0}$. For a pure initial state
$\rho_{0}=|\phi_{0}\rangle\langle\phi_{0}|$, the singular value
$\varrho_{i}=\delta_{i,1}$, then the expression of $\tau_{QSL}$ thus
recovers the unified bound of the QSLT given in Ref. \cite{13}. So
the QSLT formulated in (\ref{1}) can effectually defines the minimal
evolution time for arbitrary initial states, and also be used to
deal with the maximal speed of evolution of open quantum systems.

We consider two qubits without mutual interaction, each one
individually coupled to its own non-dissipative noisy channels.
Here, we mainly focus on the phase flip, bit flip, and bit-phase
flip channels. The dynamics of each qubit is governed by a master
equation that gives rise to a completely positive trace-preserving
map $\varepsilon(\cdot)$. In the Born-Markov approximation, the
operator-sum representation is given by
$\varepsilon(\rho_{AB})=\sum_{i,j}K^{A}_{i}K^{B}_{j}\rho_{AB}K^{B\dag}_{j}K^{A\dag}_{i}$,
with $K^{A(B)}_{i}$ are the Kraus operators that describe the noise
channels $A$ and $B$. In order to clear the QSLT of the two-qubit
initial mixed states in the above noisy channels, we take two-qubit
Bell-diagonal state as the initial state of the composite system,
described by
\begin{equation}
\rho_{0}=\frac{1}{4}(\mathbb{I}_{A}\otimes\mathbb{I}_{B}+\sum^{3}_{i=1}c_{i}\sigma^{A}_{i}\otimes\sigma^{B}_{i}),
\label{2}
\end{equation}
where $c_{i}\in\Re$ such that $0\leq|c_{i}|\leq1$ for $i=1,2,3$, and
$\mathbb{I}_{A(B)}$ is the identity operator in subspace $A(B)$. The
state in Eq. (\ref{2}) includes the Werner states
($|c_{1}|=|c_{2}|=|c_{3}|=c$) and the Bell states
($|c_{1}|=|c_{2}|=|c_{3}|=1$).

We first focus on the phase flip channel. For this channel, the
Kraus operators are given by \cite{15,17}
$K^{A}_{0}=diag(\sqrt{1-p_{A}/2},\sqrt{1-p_{A}/2})\otimes\mathbb{I}_{B}$,
$K^{A}_{1}=diag(\sqrt{p_{A}/2},\sqrt{p_{A}/2})\otimes\mathbb{I}_{B}$,
$K^{B}_{0}=\mathbb{I}_{A}{\otimes}diag(\sqrt{1-p_{B}/2},\sqrt{1-p_{B}/2})$,
$K^{B}_{1}=\mathbb{I}_{A}{\otimes}diag(\sqrt{p_{B}/2},\sqrt{p_{B}/2})$,
here, $p_{A(B)}$ is used as parametric time in channel $A(B)$, and
by considering the symmetric situation in which the decoherence rate
is equal in both channels, so $p_{A}=p_{B}$. For the initial state
of Eq. (\ref{2}), the time evolution of the total system is
\begin{eqnarray}
\rho(t)&=&\lambda^{+}_{\Psi}(t)|\Psi^{+}\rangle\langle\Psi^{+}|+\lambda^{+}_{\Phi}(t)|\Phi^{+}\rangle\langle\Phi^{+}|,\nonumber\\
&+&\lambda^{-}_{\Psi}(t)|\Psi^{-}\rangle\langle\Psi^{-}|+\lambda^{-}_{\Phi}(t)|\Phi^{-}\rangle\langle\Phi^{-}|,
\label{4}
\end{eqnarray}
where
\begin{eqnarray}
\lambda^{\pm}_{\Psi}(t)&=&(1{\pm}c_{1}(t){\mp}c_{2}(t)+c_{3}(t))/4,\nonumber\\
\lambda^{\pm}_{\Phi}(t)&=&(1{\pm}c_{1}(t){\pm}c_{2}(t)-c_{3}(t))/4,
\label{5}
\end{eqnarray}
and $|\Psi^{\pm}\rangle=(|00\rangle\pm|11\rangle)/\sqrt{2}$,
$|\Phi^{\pm}\rangle=(|01\rangle\pm|10\rangle)/\sqrt{2}$ are the four
Bell states. The time dependent coefficients in Eq. (\ref{5}) are
$c_{1}(t)=(1-p)^{2}c_{1}$, $c_{2}(t)=(1-p)^{2}c_{2}$, and
$c_{3}(t)=c_{3}$, with $p=1-\exp(-{\gamma}t)$, and $\gamma$ the
phase damping rate.

Now, we calculate the QSLT for the initial Bell-diagonal state
$\rho_{0}$ under the phase flip channel. According to the expression
(\ref{1}), we can clearly find that the Bures distance between
$\rho_{0}$ and $\rho_{\tau_{D}}$,
$B(\rho_{0},\rho_{\tau_{D}})=\frac{1}{4}(p^{2}_{\tau_{D}}-2p_{\tau_{D}})(|c_{1}|^{2}+|c_{2}|^{2})$.
Thus our main task is to calculate the singular values of $\rho_{0}$
and $\dot{\rho}_{t}$, respectively. For $\dot{\rho}_{t}$, the
singular values $\sigma_{i}$ are
\begin{eqnarray}
\sigma_{1}&=&\sigma_{2}=\frac{1}{2}|\dot{p}(1-p)(|c_{1}|+|c_{2}|)|,\nonumber\\
\sigma_{3}&=&\sigma_{4}=\frac{1}{2}|\dot{p}(1-p)(|c_{1}|-|c_{2}|)|.
\label{6}
\end{eqnarray}
While for the initial state $\rho_{0}$, the singular values
$\varrho_{i}$ depend on the region of the coefficients $c_{i}$.
Then, in the following, we divide the region of the coefficients
$c_{i}$ into four parts.

Case I: If $|c_{3}|\geq|c_{1}|\geq|c_{2}|$ in Eq. (\ref{2}), the
singular values $\varrho_{i}$ of the initial states $\rho_{0}$
satisfy
$\varrho_{1}+\varrho_{2}=\frac{1}{2}(1+|c_{3}|),\varrho_{3}+\varrho_{4}=\frac{1}{2}(1-|c_{3}|)$.
In this case,
$\sum^{4}_{i=1}\sigma_{i}\varrho_{i}=\frac{1}{2}(1-p)|\dot{p}|(|c_{1}|+|c_{2}||c_{3}|)$
is always less than
$\sqrt{\sum^{4}_{i=1}\sigma^{2}_{i}}=(1-p)|\dot{p}|\sqrt{|c_{1}|^{2}+|c_{2}|^{2}}$,
so the QSLT for the class of initial states in this case can be
obtained
$\tau^{I}_{QSL}=\frac{\tau_{D}(|c_{1}|^{2}+|c_{2}|^{2})}{|c_{1}|+|c_{2}||c_{3}|}.$

Case II: If $|c_{3}|\geq|c_{2}|\geq|c_{1}|$, we obtain
$\sum^{4}_{i=1}\sigma_{i}\varrho_{i}=\frac{1}{2}(1-p)|\dot{p}|(|c_{2}|+|c_{1}||c_{3}|)$,
the ML type bound on the QSLT is also tight for the initial states
in the case II, then we acquire
$\tau^{II}_{QSL}=\frac{\tau_{D}(|c_{1}|^{2}+|c_{2}|^{2})}{|c_{2}|+|c_{1}||c_{3}|}$.

Case III: When $|c_{1}|\geq|c_{3}|,|c_{2}|$, the singular values
$\varrho_{i}$ are given by
$\varrho_{1}+\varrho_{2}=\frac{1}{2}(1+|c_{1}|),\varrho_{3}+\varrho_{4}=\frac{1}{2}(1-|c_{1}|)$,
and
$\sum^{4}_{i=1}\sigma_{i}\varrho_{i}=\frac{1}{2}(1-p)|\dot{p}||c_{1}|(1+|c_{2}|)$
is also less than $\sqrt{\sum^{4}_{i=1}\sigma^{2}_{i}}$. The QSLT of
the initial states in case III can be written
$\tau^{III}_{QSL}=\frac{\tau_{D}(|c_{1}|^{2}+|c_{2}|^{2})}{|c_{1}|(1+|c_{2}|)}$.
The QSLT does not depend on the coefficient $c_{3}$.

Case IV: Finally, if $|c_{2}|\geq|c_{3}|,|c_{1}|$, we acquire
$\varrho_{1}+\varrho_{2}=\frac{1}{2}(1+|c_{2}|),\varrho_{3}+\varrho_{4}=\frac{1}{2}(1-|c_{2}|)$,
and
$\sum^{4}_{i=1}\sigma_{i}\varrho_{i}=\frac{1}{2}(1-p)|\dot{p}||c_{2}|(1+|c_{1}|)$.
With this, it is easy to show that the QSLT in this case yields
$\tau^{IV}_{QSL}=\frac{\tau_{D}(|c_{1}|^{2}+|c_{2}|^{2})}{|c_{2}|(1+|c_{1}|)}$
(also independent of the coefficient $c_{3}$).

Inspection of the expressions $\tau^{I}_{QSL}$, $\tau^{II}_{QSL}$,
$\tau^{III}_{QSL}$ and $\tau^{IV}_{QSL}$ shows that in all four
cases the QSLT mainly depends on the coefficients $c_{i}$ of the
initial state $\rho_{0}$. Particularly, for all states in cases
$|c_{3}|\geq|c_{2}|,|c_{1}|$ (cases I and II), the QSLT becomes
inversely proportional to $|c_{3}|$. While for all states in which
the coefficients $c_{i}$ belong to cases III and IV, the QSLT is
independent of $c_{3}$. In order to investigate the effects of
different values of the coefficients $c_{i}$ on the QSLT under phase
flip channel, we plot the QSLT of the initial Bell-diagonal states
$\rho_{0}$ as a function of $|c_{1}|$ and $|c_{2}|$ for a given
choice $|c_{3}|=0.4$, in Fig. $1$. And then, two remarkable features
can be acquired from Fig. $1$: (i) The QSLT is symmetrical with
respect to the line $|c_{1}|=|c_{2}|$. That is to say, the states
$\rho_{0}(|c_{1}|,|c_{2}|,|c_{3}|)$ and
$\rho_{0}(|c_{2}|,|c_{1}|,|c_{3}|)$ have the same quantum speed of
the evolution under phase flip channel. (ii) The QSLTs for the
initial states in cases I and II are always smaller than for those
states in cases III and IV. So we note that, under the local phase
flip channel, the Bell-diagonal states in cases I and II have
smaller minimal evolution time than those in cases III and IV.

\begin{figure}[tbh]
\includegraphics*[bb=0 0 287 246,width=7cm, clip]{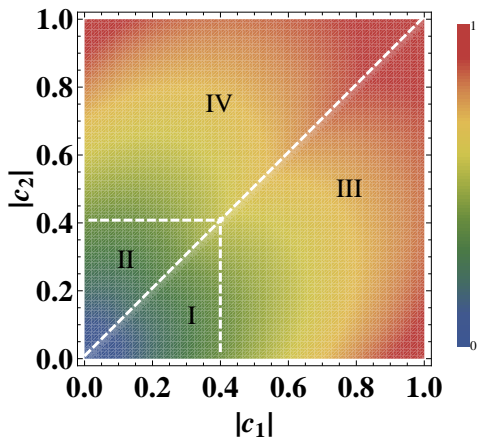}
\caption{(Color online) The QSLT under phase flip channel as a
function of the coefficients $|c_{1}|$ and $|c_{2}|$ with
$|c_{3}|=0.4$, here the driving time $\tau_{D}=1$. The dashed white
lines divide the entire range of $|c_{1}|$ and $|c_{2}|$ into four
parts: case I, $|c_{3}|\geq|c_{1}|\geq|c_{2}|$; case II,
$|c_{3}|\geq|c_{2}|\geq|c_{1}|$; case III,
$|c_{1}|\geq|c_{3}|,|c_{2}|$; and case IV,
$|c_{2}|\geq|c_{3}|,|c_{1}|$.}
\end{figure}

For the bit flip channel, the Kraus operators are \cite{15,17}
$K^{A}_{0}=diag(\sqrt{1-p/2},\sqrt{1-p/2})\otimes\mathbb{I}_{B},$
$K^{A}_{1}=\sqrt{p/2}\sigma^{A}_{x}\otimes\mathbb{I}_{B},$
$K^{B}_{0}=\mathbb{I}_{A}{\otimes}diag(\sqrt{1-p/2},\sqrt{1-p/2}),$
$K^{B}_{1}=\mathbb{I}_{A}{\otimes}\sqrt{p/2}\sigma^{B}_{x},$ The
eigenvalue spectrum of $\rho_{AB}(t)$ is also given by Eq.
(\ref{5}), where the time dependent coefficients are
$c_{1}(t)=c_{1}$, $c_{2}(t)=(1-p)^{2}c_{2}$, and
$c_{3}(t)=(1-p)^{2}c_{3}$. The QSLT of the Bell-diagonal state
$\rho_{0}$ under bit flip channel is symmetrical to that for the
phase flip channel. We just exchange the position of $c_{1}$ and
$c_{3}$ in all expressions for the phase flip channel, so the QSLT
under the bit flip channel can be obtained
\begin{eqnarray}
\tau^{B}_{QSL}= \left\{ \begin{array}{ll}
        \frac{\tau_{D}(|c_{2}|^{2}+|c_{3}|^{2})}{|c_{3}|+|c_{1}||c_{2}|}, & |c_{1}|\geq|c_{3}|\geq|c_{2}| \\[2mm]
\frac{\tau_{D}(|c_{2}|^{2}+|c_{3}|^{2})}{|c_{2}|+|c_{1}||c_{3}|}, & |c_{1}|\geq|c_{2}|\geq|c_{3}| \\[2mm]
\frac{\tau_{D}(|c_{2}|^{2}+|c_{3}|^{2})}{|c_{3}|(1+|c_{2}|)}, & |c_{3}|\geq|c_{1}|,|c_{2}| \\[2mm]
        \frac{\tau_{D}(|c_{2}|^{2}+|c_{3}|^{2})}{|c_{2}|(1+|c_{3}|)}, & |c_{2}|\geq|c_{1}|,|c_{3}| \end{array}.\right. \label{7}
\end{eqnarray}
For $|c_{3}|\geq|c_{1}|,|c_{2}|$ and $|c_{2}|\geq|c_{1}|,|c_{3}|$,
under bit flip channel, the QSLT of the initial state does not
depend on $c_{1}$. And the QSLT is symmetrical with respect to the
line $|c_{2}|=|c_{3}|$, i.e.,
$\tau^{B}_{QSL}(\rho_{0}(|c_{1}|,|c_{2}|,|c_{3}|))=\tau^{B}_{QSL}(\rho_{0}(|c_{1}|,|c_{3}|,|c_{2}|))$.
Furthermore, the Bell-diagonal states in cases
$|c_{1}|\geq|c_{3}|\geq|c_{2}|$ and $|c_{1}|\geq|c_{2}|\geq|c_{3}|$
have the smaller QSLTs than those in cases
$|c_{3}|\geq|c_{1}|,|c_{2}|$ and $|c_{2}|\geq|c_{1}|,|c_{3}|$.

Finally, for the bit-phase flip channel, the Kraus operators can be
shown \cite{15,17}
$K^{A}_{0}=diag(\sqrt{1-p/2},\sqrt{1-p/2})\otimes\mathbb{I}_{B},$
$K^{A}_{1}=\sqrt{p/2}\sigma^{A}_{y}\otimes\mathbb{I}_{B},$
$K^{B}_{0}=\mathbb{I}_{A}{\otimes}diag(\sqrt{1-p/2},\sqrt{1-p/2}),$
$K^{B}_{1}=\mathbb{I}_{A}{\otimes}\sqrt{p/2}\sigma^{B}_{y}$. In this
noise channel, the time dependent coefficients are
$c_{1}(t)=(1-p)^{2}c_{1}$, $c_{2}(t)=c_{2}$, and
$c_{3}(t)=(1-p)^{2}c_{3}$. Once more, by swapping $c_{2}$ and
$c_{3}$ in the phase flip channel, we can obtain
\begin{eqnarray}
\tau^{B-P}_{QSL}= \left\{ \begin{array}{ll}
        \frac{\tau_{D}(|c_{1}|^{2}+|c_{3}|^{2})}{|c_{3}|+|c_{2}||c_{1}|}, & |c_{2}|\geq|c_{3}|\geq|c_{1}| \\[2mm]
\frac{\tau_{D}(|c_{1}|^{2}+|c_{3}|^{2})}{|c_{1}|+|c_{2}||c_{3}|}, & |c_{2}|\geq|c_{1}|\geq|c_{3}| \\[2mm]
\frac{\tau_{D}(|c_{1}|^{2}+|c_{3}|^{2})}{|c_{1}|(1+|c_{3}|)}, & |c_{1}|\geq|c_{2}|,|c_{3}| \\[2mm]
        \frac{\tau_{D}(|c_{1}|^{2}+|c_{3}|^{2})}{|c_{3}|(1+|c_{1}|)}, & |c_{3}|\geq|c_{2}|,|c_{1}| \end{array}.\right. \label{8}
\end{eqnarray}
The fact that, for $|c_{1}|\geq|c_{2}|,|c_{3}|$ and
$|c_{3}|\geq|c_{2}|,|c_{1}|$, the coefficient $c_{2}$ would not
affect the QSLT under the bit-phase flip channel. It is also simple
to see that, for the bit-phase flip channel,
$\tau^{B-P}_{QSL}(\rho_{0}(|c_{1}|,|c_{2}|,|c_{3}|))=\tau^{B-P}_{QSL}(\rho_{0}(|c_{3}|,|c_{2}|,|c_{1}|))$.
And the QSLTs for the initial states in cases
$|c_{2}|\geq|c_{3}|\geq|c_{1}|$ and $|c_{2}|\geq|c_{1}|\geq|c_{3}|$
are always smaller than for those states in cases
$|c_{1}|\geq|c_{2}|,|c_{3}|$ and $|c_{3}|\geq|c_{2}|,|c_{1}|$.

It is worth mentioning that the QSLT of any Bell-diagonal states
under phase flip, bit flip, and bit-phase flip channels, can be
accurately calculated. This is one of our main results in this work.
Particularly, for the initial Werner state
($|c_{1}|=|c_{2}|=|c_{3}|=c$), the QSLTs under these three noise
channel are all equal to $\frac{2\tau_{D}}{1+1/c}$, which is
proportional to $c$.

\section{\textbf{{Quantum evolution speed for open dynamics process}}}

In what follows, we shall illustrate the application of the QSLT to
the quantum evolution speed of the whole dynamics process under
different decoherence channels. The QSLT can demonstrate the speed
of the dynamics evolution from a time-evolution state $\rho_{\tau}$
to another $\rho_{\tau+\tau_{D}}$ by a driving time $\tau_{D}$. It
is easy to find that the time-evolution state $\rho_{\tau}$ at any
point in time also maintains the form of Bell-diagonal state. So the
QSLT starting from an arbitrary time-evolution state $\rho_{\tau}$
can be calculated by the expressions of $\tau_{QSL}$ in Section II.
We would examine the evolution speed of the whole dynamics process
where the system starts from the two-qubit Bell-diagonal state in
Eq. (\ref{2}), under phase flip, bit flip and bit-phase flip
channels, respectively.

In the first place, under the phase flip channel,
$|c_{3}|\geq|c_{1}|,|c_{2}|$ is chosen for $\rho_{0}$. During the
dynamics evolution process, $c_{3}$ remains unchanged, so the
condition $|c_{3}|\geq|c_{1}(\tau)|,|c_{2}(\tau)|$ is fulfilled in
$\tau\in[0,\infty)$. By considering an arbitrary mixed state
$\rho_{\tau}$ evolves to another$\rho_{\tau+\tau_{D}}$ under a
driving time $\tau_{D}$, the QSLT can be calculated
\begin{eqnarray}
\tau^{P}_{QSL}= \left\{ \begin{array}{ll}
        \frac{\tau_{D}(1-p_{\tau})^{2}(|c_{1}|^{2}+|c_{2}|^{2})}{|c_{1}|+|c_{2}||c_{3}|}, & |c_{1}|\leq|c_{2}| \\[2mm]
\frac{\tau_{D}(1-p_{\tau})^{2}(|c_{1}|^{2}+|c_{2}|^{2})}{|c_{2}|+|c_{1}||c_{3}|},
& |c_{1}|\geq|c_{2}| \\[2mm]
 \end{array}.\right.\label{9}
\end{eqnarray}
From Eq. (\ref{9}), one sees immediately that, the QSLT for the
time-dependent state $\rho_{\tau}$ decays monotonically, that is to
say, in this case, the whole evolution of the two-qubit
Bell-diagonal state can exhibits a speeded-up process under the
local phase flip noise channels.

\begin{figure}[tbh]
\includegraphics*[bb=0 0 241 167,width=7cm, clip]{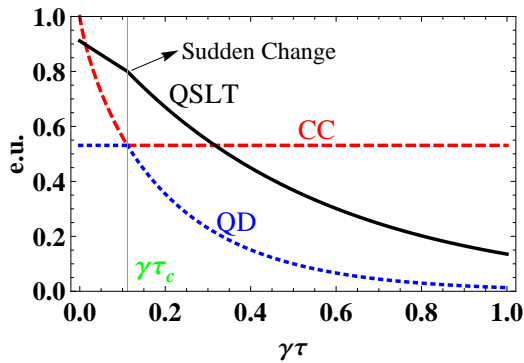}
\caption{(Color online) The QSLT (solid black line), classical
correlation (CC, dashed red line) and quantum correlation (QD,
dotted blue line) under local phase flip channels as a function of
the initial time parameter $\gamma\tau$. We choose the class of
initial Bell-diagonal states considered in Ref. \cite{16}, in which
$c_{1}=\pm1$, $c_{2}={\mp}c_{3}$, here $c_{1}=1$, $c_{2}=-0.8$, and
$c_{3}=0.8$, and the driving time $\tau_{D}=1$. On this initial
condition, the behavior of sudden transition from classical to
quantum decoherence occurs at $\tau=\tau_{c}$. The sudden change of
the decay of QSLT has been marked in this figure. }
\end{figure}

 However, if $\rho_{0}$ satisfies $|c_{1}|\geq|c_{2}|,|c_{3}|$ or
$|c_{2}|\geq|c_{1}|,|c_{3}|$, due to $c_{3}$ independence of the
evolution time, it is easy to find that, for $\tau\leq\tau_{c}$,
with
$\tau_{c}=\frac{1}{2}\ln(\frac{max\{|c_{1}|,|c_{2}|\}}{|c_{3}|})$,
the conditions $|c_{1}(\tau)|\geq|c_{2}(\tau)|,|c_{3}|$ or
$|c_{2}(\tau)|\geq|c_{1}(\tau)|,|c_{3}|$ are always true. By
considering the dynamics process $\tau\in[0,\tau_{c}]$ , we can
calculate the QSLT for the qubits to evolve from $\rho_{\tau}$ to
$\rho_{\tau+\tau_{D}}$,
\begin{eqnarray}
\tau^{P}_{QSL}= \left\{ \begin{array}{ll}
        \frac{\tau_{D}(1-p_{\tau})^{2}(|c_{1}|^{2}+|c_{2}|^{2})}{|c_{1}|[1+(1-p_{\tau})^{2}|c_{2}|]}, & |c_{1}|\geq|c_{2}|,|c_{3}| \\[2mm]
\frac{\tau_{D}(1-p_{\tau})^{2}(|c_{1}|^{2}+|c_{2}|^{2})}{|c_{2}|[1+(1-p_{\tau})^{2}|c_{1}|]},
& |c_{2}|\geq|c_{1}|,|c_{3}| \\[2mm]
 \end{array}.\right.\label{10}
\end{eqnarray}
On the other hand, for $\tau>\tau_{c}$, then
$|c_{3}|\geq|c_{1}(\tau)|\geq|c_{2}(\tau)|$ or
$|c_{3}|\geq|c_{2}(\tau)|\geq|c_{1}(\tau)|$, so when
$\tau\in(\tau_{c},\infty)$, the expressions of QSLT can be obtained
by Eq. (\ref{9}). Hence,  in the conditions
$|c_{1}|\geq|c_{2}|,|c_{3}|$ or $|c_{2}|\geq|c_{1}|,|c_{3}|$, we
note that, the decay rate of the QSLT changes suddenly at
$\tau=\tau_{c}$. So we can conclude that, in the above case, the
open system executes a speeded-up dynamics evolution process, but
the increasing rate of the evolution speed would have a sudden
change at $\tau=\tau_{c}$. Let us, finally, study the QSLT,
classical correlation  and quantum correlation (quantum discord) on
the class of initial states for which $c_{1}=\pm1$ and
$c_{2}={\mp}c_{3}$, with $|c_{3}|<1$, in Fig. $2$. It is worth
noting that, as noted in Ref. \cite{16}, for this initial condition,
the behavior of sudden transition from classical to quantum
decoherence can occur at $\tau=\tau_{c}$. And at this time point,
the decay rate of the QSLT can also change suddenly. Remarkably, the
attenuations of the QSLT and classical correlation are synchronous
in time until $\tau=\tau_{c}$. However, for $\tau>\tau_{c}$, the
synchronous attenuation behavior occurs between the QSLT and quantum
discord. This means that the speed of the dynamics evolution from a
time-evolution state $\rho_{\tau}$ to another $\rho_{\tau+\tau_{D}}$
can be clearly signatured by the classical correlation of the
time-evolution state $\rho_{\tau}$ in the case $\tau<\tau_{c}$.
While for $\tau>\tau_{c}$, we can use the quantum discord of
$\rho_{\tau}$ to describe the speed of dynamics process,
qualitatively. This is a newly noticed phenomenon.

Finally, when we consider the bit flip and the bit-phase flip
channels, the behaviors of the evolution speed of the whole dynamics
process under the phase flip channel described above can also occur
under other conditions on the $c_{i}$ in state (\ref{2}). It is
simple to see that, for the bit flip and bit-phase flip channels,
the QSLT starting from an arbitrary time-evolution state
$\rho_{\tau}$ to $\rho_{\tau+\tau_{D}}$ can be obtained with $c_{1}$
and $c_{2}$ replacing $c_{3}$, respectively.

\section{\textbf{{Conclusion}}}
In summary, based on the unified bound of QSLT for arbitrary mixed
states, we have calculated the expressions of the QSLT to
characterize the speed of evolution for an open composite quantum
system. In particular, for the initial two-qubit Bell-diagonal
state, we have identified four different expressions of the QSLT
under decoherence, which depend on the coefficients of the initial
composite state and on the noise channel. Moreover, by studying the
bound for the arbitrary time-dependent states started from a class
of two-qubit states cited in Ref. \cite{16}, a remarkable feature
has been demonstrated that the decay rate of the QSLT can change
suddenly at a critical time $\tau_{c}$, at this moment the behavior
of sudden transition from classical to quantum decoherence also
occurs. This evolution speed of the dynamics process plays an
essential role in the understanding of the classical-to-quantum
decoherence transition. Our results may be of interests in exploring
the speed of quantum computation and quantum information processing
in the presence of noise.

\section{\textbf{{Acknowledgement}}}
This work was supported by ¡°973¡± program under grant No.
2010CB922904, the National Natural Science Foundation of China under
grant Nos. 11175248, 61178012, 11304179, 11247240, the Specialized
Research Fund for the Doctoral Program of Higher Education under
grant Nos. 20123705120002, 20133705110001 and the Provincial Natural
Science Foundation of Shandong under grant No. ZR2012FQ024.

\end{document}